\documentclass[aps,prb,twocolumn,showpacs,byrevtex]{revtex4}

\usepackage{times,mathptm}
\usepackage[dvips]{graphicx,color}

\begin{document}

\title{Contrasting interedge superexchange interactions of graphene nanoribbons embedded in $h$-BN and graphane}
\author{Sun-Woo Kim,$^{1}$ Hyun-Jung Kim,$^{1}$ Jin-Ho Choi,$^{2,1}$ Ralph H. Scheicher,$^{3}$ and Jun-Hyung Cho$^{1}$$^{*}$}
\affiliation{$^{1}$Department of Physics and Research Institute for Natural Sciences, Hanyang University, 17 Haengdang-Dong, Seongdong-Ku, Seoul 133-791, Korea\\
$^{2}$Hefei National Laboratory for Physical Sciences at the Microscale and Department of Physics, University of Science and Technology of China, 96 JinZhai Road, Hefei, Anhui 230026, China\\
$^{3}$Division of Materials Theory, Department of Physics and Astronomy, {\AA}ngstr{\"o}m Laboratory, Uppsala University, Box 516, SE-751 20, Uppsala, Sweden}
\date{\today}
\begin{abstract}
Based on first-principles density-functional theory calculations, we present a comparative study of the electronic structures of ultranarrow zigzag graphene nanoribbons (ZGNRs) embedded in hexagonal boron nitride (BN) sheet and fully hydrogenated graphene (graphane) as a function of their width $N$ (the number of zigzag C chains composing the ZGNRs). We find that ZGNRs/BN have the nonmagnetic ground state except at $N$ = 5 and 6 with a weakly stabilized half-semimetallic state, whereas ZGNRs/graphane with $N$ ${\ge}$ 2 exhibit a strong antiferromagnetic coupling between ferromagnetically ordered edge states on each edge. It is revealed that the disparate magnetic properties of the two classes of ZGNRs are attributed to the contrasting interedge superexchange interactions arising from different interface structures: i.e., the asymmetric interface structure of ZGNRs/BN gives a relatively short-range and weak superexchange interaction between the two inequivalent edge states, while the symmetric interface structure of ZGNRs/graphane gives a long-range, strong interedge superexchange interaction.
\end{abstract}

\maketitle

Graphene nanoribbons (GNRs) have been regarded as one of the most important classes of carbon-based nanomaterials due to their unique electronic and magnetic properties.~\cite{Fujita, Nakada, YWSon1, YWSon2} Fabrication of GNRs with different widths and edges has been achieved by lithographic patterning,~\cite{MYHan} bottom-up fabrication,~\cite{Cai} and chemical unzipping of carbon nanotubes.~\cite{Kosynkin,Jiao,Tao} It is known that the electronic and magnetic properties of GNRs vary with the ribbon width and the edge geometry,~\cite{Fujita, Nakada, YWSon1, YWSon2} thereby being utilized to design novel electronic and spintronic devices. For instance, the band gap of GNRs depends on the ribbon width,~\cite{YWSon1} and the zigzag-edged graphene nanoribbons (ZGNRs) have peculiar localized electronic states at both edges while the GNRs with armchair edges do not have such localized edge states.~\cite{Fujita, Nakada} Here, the localized edge states of ZGNRs are ferromagnetically ordered at each edge with an opposite spin orientation, forming an antiferromagnetic (AFM) spin order. Interestingly, it was predicted that such an AFM ordered ZGNR can have a half-metallic property if in-plane homogeneous electric field is applied across the edges of the ZGNR.~\cite{YWSon2} However, the applied electric field is practically too high to realize half-metallic ZGNRs,~\cite{YWSon2,Kan1} and therefore various alternative approaches~\cite{Hod,Gunlycke,Kan2,Wu} have been proposed. Most of the alternatives focus essentially on the same conceptual basis that the half-metallicity of ZGNRs can be enabled by the modification of edge states: e.g., the edge modification of ZGNRs with two different functional groups can produce the half-metallic property even in the absence of an external electric field.~\cite{Kan2,Wu} Such an asymmetric edge modification can also be achieved when ZGNRs are embedded in a hexagonal boron nitride ($h$-BN) sheet.~\cite{Ding,Pruneda,Jungthawan,Liu1,Liu2} These embedded ZGNRs (hereafter designated as ZGNRs/BN) have a C$-$B interface at one edge and a C$-$N interface at the opposite edge [see Fig. 1(a)]. A number of density functional theory (DFT) calculations reported the presence of half-metallicity or half-semimetallicity in ZGNRs/BN.~\cite{Ding,Pruneda,Jungthawan} On the experimental side, the fabrication of graphene-BN hybrid structures was recently reported, with graphene strips as narrow as tens of nanometers.~\cite{Liu1,Liu2} In a different way, the ultranarrow ZGNRs can be fabricated~\cite{Wang} by removing hydrogen atoms from a fully hydrogenated graphene (viz. graphane):~\cite{Sofo} see Fig. 1(b).
\begin{figure}[ht]
\centering{ \includegraphics[width=7cm]{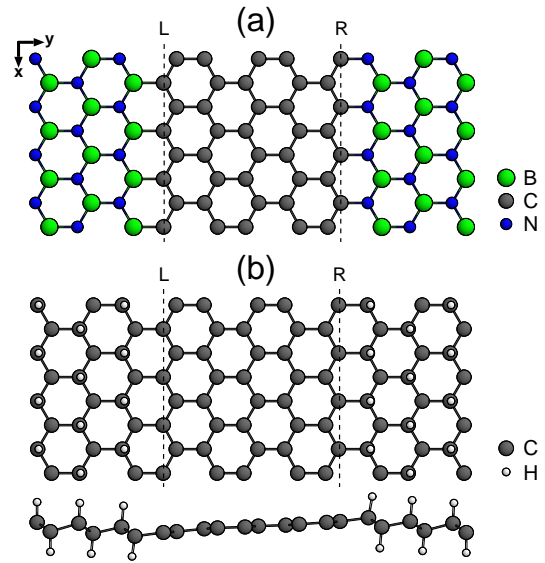} }
\caption{(Color online) Top view and side view of the optimized structures of (a) the ZGNR/BN and (b) the ZGNR/graphane with $N$ = 5. The $x$ and $y$ axes are taken to be parallel and perpendicular to the edges of ZGNR, respectively. L and R represent the left and right edges, respectively.}
\end{figure}
\begin{figure*}[ht]
\centering{ \includegraphics[width=14cm]{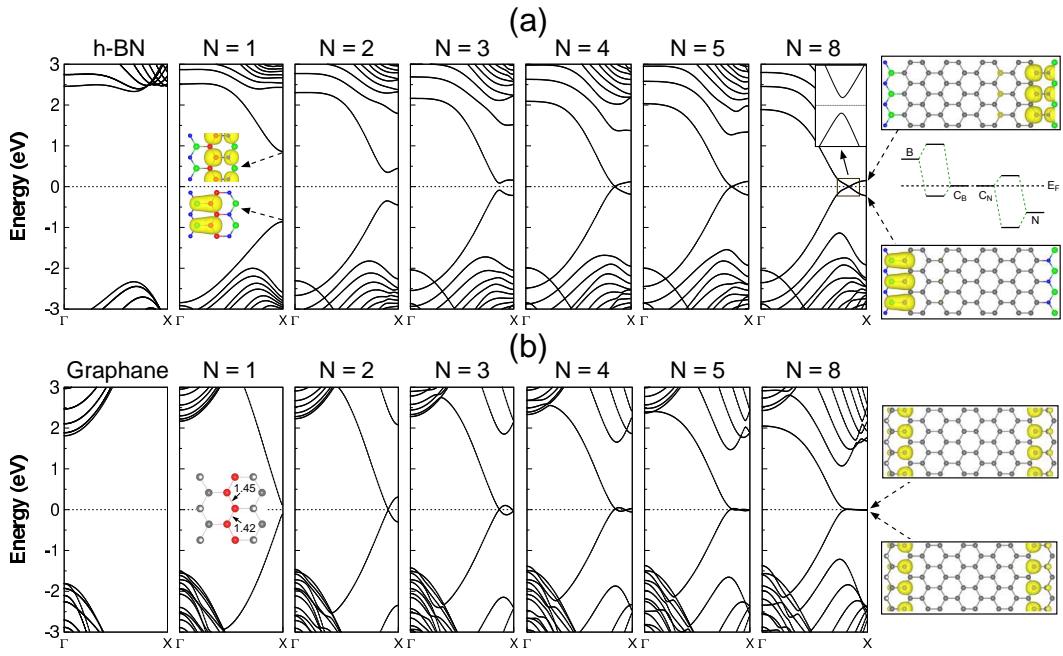} }
\caption{(Color online) Calculated band structures of (a) ZGNRs/BN and (b) ZGNRs/graphane as a function of $N$. The results for the $h$-BN sheet and graphane are also given. The charge character of edge states is shown with an isosurface of 0.01 electrons/{\AA}$^3$. In the schematic diagram, ${\rm C}_{\rm B}$ (${\rm C}_{\rm N}$) indicates the edge C atom bonding to a B (N) edge atom. The energy zero represents the Fermi level. The direction of $\Gamma$ $-$ X line is parallel to the edges. For distinction, C atoms composing the ZGNR with $N$ = 1 are drawn with circles in different brightness.}
\end{figure*}
Such embedded ZGNRs (hereafter designated as ZGNRs/graphane) were predicted to exhibit the insulating AFM ground state,~\cite{Singh,JHLee,HJKim} similar to isolated ZGNRs.~\cite{YWSon2} Note that ZGNRs/graphane have the symmetric interface structure with an identical C$-$CH interface on both edges, differing from the asymmetric interface structure of ZGNRs/BN. Therefore, it is very interesting to explore the roles of interface structure in determining the drastically different electronic and magnetic properties of ZGNRs/BN and ZGNRs/graphane.

In this paper, we perform first-principles DFT calculations to investigate the electronic structures of ultranarrow ZGNRs/BN and ZGNRs/graphane as a function of $N$ from 1 to 8. We find that, in contrast to ZGNRs/graphane where the nonmagnetic (NM) ground state at $N$ = 1 is converted to the AFM ground state for $N$ ${\geq}$ 2, ZGNRs/BN have the NM ground state except at $N$ = 5 and 6 with the half-semimetallic state. Such different behaviors of the ZGNRs embedded in $h$-BN sheet and graphane can be traced to the contrasting features of edge states due to their different interface structures. The asymmetric interface structure of ZGNRs/BN produces the inequivalent edge states originating from the B$-$C and N$-$C interfaces, giving rise to a relatively short-range and weak superexchange interaction between the two edge states. On the other hand, the symmetric interface structure of ZGNRs/graphane produces the identical edge states with partially flat bands, leading not only to a magnetic instability due to the enhanced density of states (DOS) at the Fermi level ($E_F$) but also to a long-range, strong interedge superexchange interaction.

The present first-principles DFT calculations were performed using the Fritz-Haber-Institute $ab$-initio molecular simulations (FHI-aims) code~\cite{Aims} for an accurate, all-electron description based on numeric atom-centered orbitals, with ``tight" computational settings and accurate tier-2 basis sets. For the exchange-correlation energy, we employed the generalized gradient approximation functional of Perdew-Burke-Ernzerhof.~\cite{PBE} The embedded ZGNRs were simulated using a periodic supercell with a constant in-plane unit cell length of ${\sim}$46 {\AA} (changing the width of BN or graphane from ${\sim}$26 to ${\sim}$41 {\AA} with respect to $N$) and a vacuum spacing of ${\sim}$30 {\AA} between the periodic sheets. For the Brillouin zone integration, we used 128${\times}$1${\times}$1 ${\bf k}$-points in the surface Brillouin zone. All of the atoms were allowed to relax along the calculated forces until all the residual force components are less than 0.02 eV/{\AA}. The optimized lattice constant and band gap of $h$-BN sheet (graphane) are found to be 2.51 (2.54) {\AA} and 4.66 (3.62) eV, respectively, in good agreement with previous DFT calculations.~\cite{Sofo,Topsakal}

\begin{table*}[ht]
 \caption{
  Calculated energy difference (in meV/unit cell) between the NM and half-semimetallic (HS) or AFM configurations for ZGNRs/BN and ZGNRs/graphane as a function of $N$. The band gap in each system is also given in the unit of eV.
 }
 \begin{ruledtabular}
 \begin{tabular}{lccccccccc}
  &  &    $N$ $=$ 1   &  $N$ $=$ 2    &  $N$ $=$ 3 & $N$ $=$ 4 & $N$ $=$ 5 & $N$ $=$ 6 & $N$ $=$ 7 & $N$ $=$ 8 \\  \hline
  ZGNRs/BN & ground state   &  NM  &  NM  &  NM  &  NM  &  HS  &  HS  &  NM  &  NM  \\
  & ${\Delta}E_{\rm NM-HS}$ &  $-$  & $-$   & $-$   & $-$   &  1.7  &  1.5  & $-$   &  $-$  \\
  & $E_g$  &   1.711   &   0.701   &   0.176   &   0.020   &     0.023       &      0.002      &   0.005   &  0.010  \\
 ZGNRs/graphane & ground state   &  NM  &  AFM  &  AFM  &  AFM  &  AFM  &  AFM  &  AFM  &  AFM  \\
    & ${\Delta}E_{\rm NM-AFM}$ &  $-$  &  8.0  &  32.6  &  50.3  &  59.5  &  64.5  &  69.1  &  72.7     \\
  & $E_g$   &  0.217   &   0.410   &   0.532   &   0.533   &      0.509      &     0.479       &   0.450   &  0.422  \\
 \end{tabular}
 \end{ruledtabular}
 \end{table*}

We begin to determine the atomic and electronic structures of ZGNRs/BN and ZGNRs/graphane using spin-unpolarized calculations. The calculated band structures of ZGNRs/BN and ZGNRS/graphane are displayed as a function of $N$ in Fig. 2(a) and 2(b), respectively, together with those of $h$-BN sheet and graphane. For the ZGNR/graphane with $N$ = 1, we obtain a bond-alternated structure with two different C$-$C bond lengths, $d_{\rm C-C}$ = 1.42 and 1.45 {\AA} [see the inset of Fig. 2(b)], indicating a Peierls distortion of ${\Delta}d$ = ${\pm}$0.015 ${\AA}$. This Peierls distortion accompanies a band-gap opening of 0.22 eV between the ${\pi}$ and ${\pi}^*$ bands. On the other hand, for the ZGNR/BN with $N$ = 1, such a bond alternation does not occur with an equal C$-$C bond length of $d_{\rm C-C}$ = 1.43 {\AA}, and the charge character of the ${\pi}$ (${\pi}^*$) state at the $X$ point represents the hybridization between C and B (N) atoms [see the inset of Fig. 2(a)], giving rise to a large band-gap opening of 1.71 eV. As $N$ increases, the C$-$C bond lengths in ZGNRs/BN and ZGNRs/graphane are close to each other as 1.44 and 1.45 {\AA}, respectively. Figure 2(a) shows that, as $N$ of ZGNRs/BN increases, the band gap ($E_g$) decreases, almost being closed from $N$ = 4. However, we note that the breaking of the sublattice symmetry in ZGNRs/BN, due to their asymmetric interface structure, avoids the crossing of $\pi$ and $\pi^{*}$ bands at $E_F$ [see the inset of $N$ = 8 in Fig. 2(a)]. On the other hand, for ZGNRs/graphane with identical edge interfaces, the ${\pi}$ and ${\pi}^*$ bands cross the Fermi level with increasing $N$, formimg a partially two-fold degenerate flat band at a sufficiently wider width [see Fig. 2(b)].

It is noteworthy that the charge characters of $\pi$ and $\pi^{*}$ states in ZGNRs/BN represent asymmetric edge states localized at the C$-$B and C$-$N interfaces, respectively [see the inset for $N$ = 8 in Fig. 2(a)]. Here, the schematic diagram of frontier orbital interactions shows that the highest occupied ${\pi}$ and lowest unoccupied ${\pi}^*$ states are characterized as the C$-$B bonding and C$-$N antibonding orbitals, respectively. On the other hand, the ZGNR/graphane with $N$ = 8 shows symmetric $\pi$ and $\pi^{*}$ edge states localized at both sides [see the inset for $N$ = 8 in Fig. 2(b)]. These different features of edge states between ZGNRs/BN and ZGNRs/graphane may influence the range and strength of the interaction between two edges. In order to compare the effects of the interedge interaction on the half-semimetallicity of ZGNRs/BN and the AFM order of ZGNRs/graphane, we perform spin-polarized calculations for the two systems as a function of $N$. It is known that the electric field created by different electrostatic potentials at the C$-$B and C$-$N interfaces is associated with half-semimetallicity in ZGNRs/BN,~\cite{Pruneda} while the flat-band-like character in the edge states of ZGNRs/graphane induces a magnetic instability due to the enhanced DOS at $E_F$.~\cite{JHLee} The calculated stabilization energies of half-semimetallicity and AFM order in ZGNRs/BN and ZGNRS/graphane relative to the corresponding NM configuration are given as a function of $N$ in Table I. In ZGNRs/BN, we obtain the NM ground state for $N$ ${\leq}$ 4 and $N$ ${\geq}$ 7, while the half-semimetallic ground state at $N$ = 5 and 6. This trend showing that wide-(or extremely narrow) and intermediate-width ZGNRs are stabilized as the NM and half-semimetallic configurations, respectively, is consistent with previous DFT study.~\cite{Pruneda} Note that the half-semimetallic configuration at $N$ = 5 and 6 is only a few meV lower in energy than the corresponding NM configuration (see Table I), indicating that the interedge interaction producing half-semimetallicity in ZGNRs/BN is very weak. On the other hand, in ZGNRs/graphane, we obtain the AFM ground state for $N$ ${\geq}$ 2, where the total-energy difference ${\Delta}E_{\rm NM-AFM}$ between the NM and AFM configurations monotonically increases as $N$ increases, reaching ${\sim}$73 meV at $N$ = 8 (see Table I). These results obviously indicate that the interedge interaction in ZGNRs/graphane is long-range and strong compared to that in ZGNRs/BN.
\begin{figure}[ht]
\centering{ \includegraphics[width=7cm]{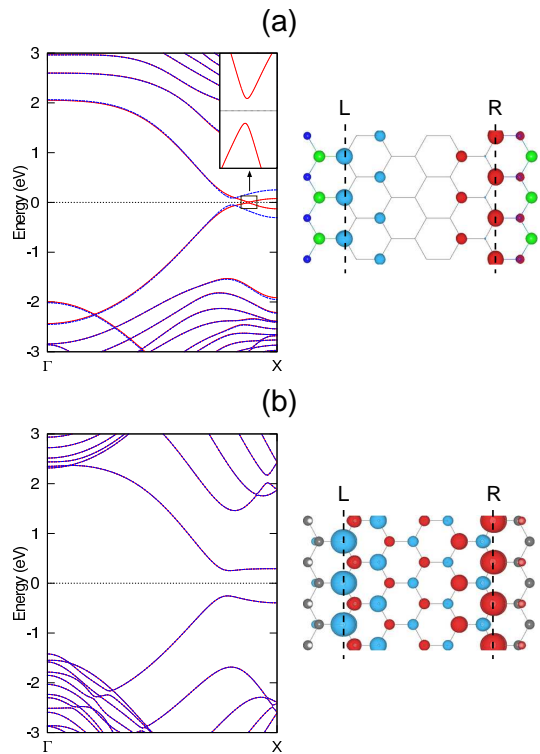} }
\caption{(Color online) Calculated AFM band structures of (a) ZGNR/BN and (b) ZGNR/graphane for $N$ = 5. The energy zero represents the Fermi level. The spin densities of ZGNR/BN and ZGNR/graphane are also given. The inset in (a) magnifies the band gap of the spin-down bands. The spin densities are drawn with an isosurface of 0.02 ($-$0.02) electrons/{\AA}$^3$.
}
\end{figure}
It is remarkable that the geometric symmetry of two edges in ZGNRs embedded in either $h$-BN sheet or graphane plays an important role in determining the range and strength of interedge interaction.

Figure 3(a) and 3(b) show the comparison of the spin-polarized band structures of the ZGNR/BN and ZGNR/graphane with $N$ = 5. In the former system, the spin-up and spin-down bands open a gap of 0.14 and 0.02 eV [see Fig. 3(a)], respectively. These values of gap opening are much smaller compared to the ZGNR/graphane system where the spin-up and spin-down bands open an identical band gap of 0.51 eV [see Fig. 3(b)]. Here, the much smaller band gap in the half-semimetallic ZGNR/BN compared to the AFM ZGNR/graphane gives rise to a much smaller value of ${\Delta}E_{\rm NM-HS}$ = 1.7 meV than ${\Delta}E_{\rm NM-AFM}$ = 59.5 meV (see Table I). It is interesting to notice that there is a subtle difference of the spin characters between the half-semimetallic ZGNR/BN and the AFM ZGNR/graphane. As shown in Fig. 3(a) and 3(b), the spin density of the former system is relatively well-localized around the two edges, whereas that of the latter system shows some extension up to the middle of the ribbon. This reflects relatively short-range (weak) versus long-range (strong) interedge spin-spin interactions in ZGNRs/BN and ZGNRs/graphane.

To understand the microscopic mechanism for the half-semimetallicity and AFM order in ZGNRs/BN and ZGNRs/graphane, we plot, in Fig. 4(a) and 4(b), the spin-polarized local DOS projected onto the two edge C atoms [in the left (L) or right (R) edge site in Fig. 1] together with their spin characters. For the ZGNR/BN with $N$ = 5, it is seen that the occupied (unoccupied) spin-up and spin-down edge states are localized at the L (R) edge [see Fig. 4(a)]. However, for the ZGNR/graphane with $N$ = 5, the occupied (unoccupied) spin-up and spin-down edge states are localized at the L (R) and R (L) edges, respectively [see Fig. 4(b)]. Since electronic states with the same spin direction can hybridize with each other, the hybridization occurs between the occupied and unoccupied spin-up or spin-down states localized at the L and R (or R and L) edges. This kind of exchange interaction between the occupied and unoccupied states is characterized as a superexchange mechanism.~\cite{Goodenough,Kanamori,Sato} Such an interedge superexchange interaction leads to a relatively long-range, strong interedge spin-spin interaction in ZGNRs/graphane, thereby giving rise to a large energy gain in ${\Delta}E_{\rm NM-AFM}$ as well as a large gap opening, as shown in Table I.

\begin{figure}[ht]
\centering{ \includegraphics[width=7.7cm]{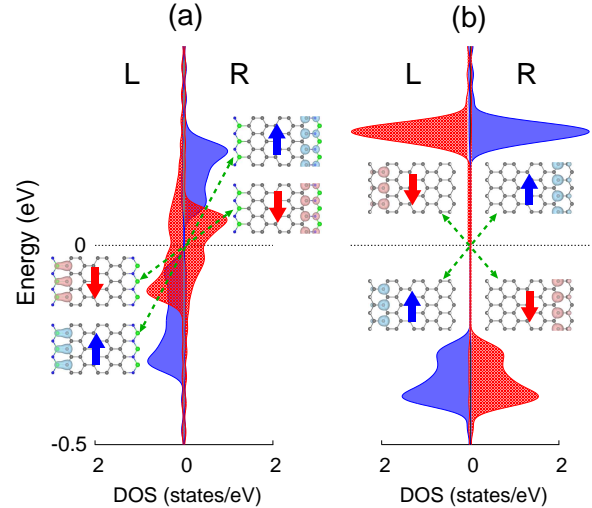} }
\caption{(Color online) The spin-polarized local DOS projected onto the two edge C atoms [in the left (L) or right (R) edge site in Fig. 1] of (a) ZGNR/BN and (b) ZGNR/graphane with $N$ = 5. The energy zero represents Fermi level. The charge characters of the spin-up and spin-down states for the occupied and the unoccupied band are taken at the X point with an isosurface of 0.04 ($-$0.04) electrons/{\AA}$^3$}
\end{figure}

In summary, using first-principles DFT calculations, we have performed a comparative study of the electronic structures of ultranarrow ZGNRs/BN and ZGNRs/graphane. Such embedded ZGNRs in $h$-BN sheet and graphane are found to exhibit drastically different electronic characteristics. Unlike ZGNRs/graphane, whose NM configuration exhibits partially flat bands at $E_F$ as $N$ increases, ZGNRs/BN do not have such a flat-band-like character. Consequently, the former ZGNRs show a magnetic instability due to the enhanced DOS at $E_F$, whereas the latter ZGNRs preserve the NM ground state except at $N$ = 5 and 6 with a half-semimetallic state. We revealed that the disparate magnetic properties of the two classes of ZGNRs can be traced to the different features of their interface structures: i.e., unlike the symmetric interface structure of ZGNRs/graphane, the asymmetric interface structure of ZGNRs/BN produces the inequivalent edge states on both sides of the nanoribbons, giving rise to a relatively short-range, weak interedge superexchange interaction. The resulting different electronic and magnetic properties of ZGNRs/BN and ZGNRs/graphane may be utilized for the application of nano-scale electronic devices such as conducting wires or field-effect transistors.

\noindent {\bf Acknowledgement.}
This work was supported by the National Research Foundation of Korea (NRF) grant funded by the Korea Government (MSIP) (Grant No. 2014M2B2A9032247) and the Korea$-$Sweden Research Cooperation Programme of the NRF (Grant No. 2011-0031286) and the Swedish Foundation for International Cooperation in Research and Higher Education (STINT, Grant No. 2011/036). R.H.S. acknowledges support from the Swedish Research Council (VR, Grant No. 621-2009-3628). The calculations were performed by KISTI supercomputing center through the strategic support program (KSC-2014-C3-049) for the supercomputing application research.

\noindent $^{*}$Corresponding author: chojh@hanyang.ac.kr

\end{document}